\documentstyle[12pt]{article}
\begin{document}
\begin{titlepage}

\pagestyle{empty}

\begin{flushright}
{\footnotesize Brown-HET-1034\\

March 1996}
\end{flushright}

\vskip 1.0cm

\begin{center}
{\Large \bf COSMOLOGIES WITH PHOTON CREATION AND THE 3K RELIC RADIATION SPECTRUM}

\vskip 1cm
J. A. S. Lima$^{1,2}$ 
\end{center}
\vskip 0.5cm
\begin{quote}

$^1$ Departamento de F\'{\i}sica Te\'orica e Experimental,
     Universidade \\ 
$^{ }$ $^{ }$ Federal do Rio Grande do Norte, 
     59072 - 970, Natal, RN, Brazil

{\small $^2$ Physics Department, Brown University, 
Providence, RI 02912, USA

({submitted to {\it Gravity Research Foundation Essay Competition}})}
\end{quote} 

\vskip 1.0cm

\centerline{\bf SUMMARY}
\bigskip

\noindent A new Planckian distribution for
cosmologies with photon creation is 
derived using 
thermodynamics and semiclassical 
considerations. This    
spectrum is preserved during the 
evolution of the universe and compatible 
with the present spectral 
shape of the cosmic microwave background 
radiation(CMBR). Accordingly, the widely spread 
feeling that 
cosmologies with 
continuous photon creation are definitely
ruled out by the COBE 
limits on deviation of the CMBR 
spectrum 
from blackbody shape should be 
reconsidered. It is argued that a crucial test  
for this kind of cosmologies is provided by 
measurements of the 
CMBR temperature at high redshifts. For 
a given 
redshift $z$ greater than zero, the 
temperature is 
smaller than the one predicted by the 
standard FRW model. 

\end{titlepage}

\newpage
\pagestyle{plain}
\baselineskip 0.75cm
\newpage

The problem investigated here may be stated in a very 
broad and simple way: If photons are continuously created during the evolution of the universe, 
under which conditions may this process be compatible with 
the present blackbody nature of the cosmic microwave 
background radiation(CMBR)? This question  was discussed long ago by H\"onl and Dehen\cite{HD 68} to rule out the gravitational theory of 
Jordan\footnote{The criticism of 
H\"onl and Dehen was subsequently accepted by P. Jordan\cite{JO 68}.} and, in a more general framework, by Steigman\cite{GS 78}. In the latter paper was concluded that any photon nonconserving cosmology 
is in conflict with the observed spectral shape of the CMBR. As a 
consequence,  the interest in cosmologies based 
on continuous photon creation, like Dirac cosmologies, G-variable models of Canuto and coworkers,
matter creation cosmologies of Hoyle and Narlikar and others\cite{MC 76} 
perceptively declined in the literature. 

The argument of Steigman may briefly be 
restated as follows: Consider an arbitrary 
spectrum of photons whose number and energy densities 
are, respectively, $n_r \sim T^
{3}$ and $\rho_r \sim T^{4}$ and let $N_r(t)$, the 
instantaneous comoving number of 
photons. Since $N_r=n_rR^{3}$, where 
R(t) is the scale factor of a FRW 
cosmology, it follows
that 
\begin{equation}
\label{eq:AF}
N_r(t)^{-\frac {1}{3}}TR = const  \quad .  
\end{equation}
On the other hand, photons in FRW geometries redshift 
away obeying $\nu \sim R^{-1}$ so that (1) may be rewritten as 
\begin{equation}
\label{eq:LT}
[N_r(t)]^
{-\frac {1}{3}}\frac {T}{\nu} = const  \quad .
\end{equation}
Therefore, the ratio $\frac {T}{\nu}$ will be an invariant in the 
course of the expansion, thereby preserving 
Planck's distribution, only if $N_r(t)=const$. In 
Steigman's words: ``{\it Unless the
number of photons in a comoving volume 
is conserved, a 
blackbody distribution 
is destroyed as the universe 
evolves }''. Naturally, the same 
criticism also holds for modern 
theories with photon 
production like decaying vacuum 
cosmologies\cite{FAF 87} or irreversible 
matter creation  at the expenses 
of the gravitational 
field\cite{PR 89,CLW 92}.

In the post-COBE(and Hubble Space Telescope) era, the 
importance of the question posed in the first 
paragraph is more easily recognized than in 
the seventies. As discussed very 
recently, the latest measurements of the Hubble 
parameter\cite{FR 95} points to an intrinsic 
fragility of the standard (photon conserving) FRW 
cosmology, in such a way that models 
without cosmological constant seems to be effectively 
ruled out\cite{KT 95}. The dillem is quite obvious. The increasing 
difficulties 
of the standard model strongly suggest that it will 
have given way to alternative Big Bang 
cosmologies. However, the finely adjusted 
blackbody nature of the CMBR 
(from COBE measurements) coupled 
with the criticism of Steigman, works like
a Damocles sword pending on the foundations of any 
cosmology endowed with continuous photon creation.  

In this Essay, I will reanalyse this question  
working in an extended 
framework. First, a formula for blackbody 
radiation when photon creation 
takes place ``adiabatically'' will be derived. This 
terminology  will be further justified 
though, for a moment, it will be
employed only to label the hypotheses assumed  by 
Steigman, namely that some 
equilibrium relations are preserved during the creation process. Second,
a crucial test for photon creation cosmologies will
also be suggested. As we shall see, the 
new spectrum is 
both preserved under expansion of the 
universe and consistent with 
the present spectral shape of the CMBR radiation.
\vskip 0.3cm
\noindent{\bf The``Adiabatic'' Blackbody Spectrum}
\vskip 0.3cm

If one 
compresses or expands a hollow 
cavity containing blackbody radiation, in such a way that 
photons (due to some unspecified 
creation process) are ``adiabatically''
added in it, then for each 
wave component one may write

\begin{equation}
N_r(t)^{-\frac {1}{3}}{\lambda}T = const.  \quad .
\end{equation}
This quantity plays the role of a generalized 
``adiabatic'' invariant in the sense of Ehrenfest\cite{PE 17}. When 
$N_r(t)$ is constant, the usual adiabatic invariant for an 
expanding blackbody radiation is recovered.

Now, let $T_{1}$ be the temperature at the instant $t=t_1$, and 
focus our attention 
on the band
$\Delta \lambda _{1}$(centered on $\lambda _{1}$) whose energy
density is $\rho _{T_{1}}(\lambda _{1})\Delta \lambda _{1}$. At a 
subsequent time $t=t_2$, when $T_{1}$ changed to $T_{2}$, the energy 
of the band changed to
$\rho _{T_{2}}(\lambda _{2})\Delta \lambda _{2}$ and, according 
to (3), $\Delta \lambda _{1}$ and $\Delta \lambda _{2}$ are 
related by
  
\begin{equation}
\label{DELTL}
{\Delta \lambda _{2} \over \Delta \lambda _{1}} = 
(\frac {N_r(t_2)}{N_r(t_1)})^
{\frac {1}{3}} {T_{1} \over T_{2}} \quad,
\end{equation}
where $N_r(t_1)$, $N_r(t_2)$ are, respectively, the 
net number of photons 
at times $t_1$ and $t_2$. 
Like Steigman, let us now assume that some 
thermodynamic equilibrium relations are preserved
(``adiabatic'' photon creation). In this 
case, since distinct bands do not interact 
 
\begin{equation}
\label{RODELT}
	{\rho _{T_{2}}(\lambda _{2})\Delta \lambda _{2} \over \rho
	_{T_{1}}(\lambda _{1})\Delta \lambda _{1}}  = ({T_{2} \over
	T_{1}})^{4} \quad.
\end{equation} 
By combining (5) with (3)
and using (4),
we obtain for an arbitrary component
$\rho_{T}(\lambda)\lambda ^{5} = const N_r(t)^{\frac {4}{3}}$. In the Planckian case ($N_r(t)=const$), this
expression reduces to
$\rho _{T} (\lambda )\lambda ^{5} = const$, as it should be. Without loss
of generality, taking into account
(3), this result may be rewritten as (we have 
normalized $N_r(t)$ by its
value $N_{or}$ without photon creation)
 
\begin{equation}
\rho_{T} (\lambda) = (\frac {N_r(t)}{N_{or}})^{\frac{4}{3}}
\lambda ^{-5}
\phi((\frac {N_r(t)}{N_{or}})^{-\frac{1}{3}}\lambda T)    \quad,
\end{equation}
where $\phi$ is an arbitrary function of its argument. In terms of
frequency, since $\rho _{T}(\nu)d\nu = \rho _{T}(\lambda )\mid {d\nu
\over d\lambda } \mid d\lambda$, it follows that

\begin{equation}
	\rho_{T} (\nu ) = \alpha (\frac {N_r(t)}{N_{or}})^{\frac {4}{3}} \nu ^{3}
\phi ((\frac {N_r(t)}{N_{or}})^{-\frac {1}{3}}{T\over \nu }) \quad,
\end{equation}
where $\alpha$ is a dimensional constant. The above equation is the 
generalized form of Wien's 
law. It reduces to the standard Wien law
when the number of photons is conserved\cite{MP 14}.

Next, following the arguments originally used by Einstein\cite{AE 17}, the complete distribution it will be derived. Consider now an 
atomic or 
molecular gas, the particles of which can
exist in a number of discrete energy levels $E_{n}=1,2,...$etc, in thermal
equilibrium with the radiation at temperature $T$. The probability
that an atom is in the energy level $E_{n}$ is 
given by the Boltzmann factor, $p_{n}e^{-{E_{n}\over kT}}$, 
where $p_{n}$ is the statistical weight of the nth quantum state. In such a system there exist three kinds of transition processes,
namely: absorption, spontaneous and stimulated emission; which are charactherized, respectively, by the coefficient $B^m_n$, $A^{n}_{m}$ and $B^{n}_{m}$. From equilibrium condition\cite{AE 17}
\begin{equation}\label{pnpm}
	p_{n}e^{-E_{n}/kT} B^{m}_{n}\rho _{T}(\nu )=p_{m}e^{-E_{m}/kT}
	(B^{n}_{m}\rho _{T}(\nu )+A^{n}_{m}) \quad,
\end{equation}
and solving for the energy density 
\begin{equation}\label{forend}
	\rho _{T}(\nu ) = {{p_{m}\over p_{n}} {A^{n}_{m}\over B^{m}_{n}} \over
			 e^{E_{m}-E_{n}\over kT} - {p_{m}\over p_{n}}
			 {B^{n}_{m}\over B^{m}_{n}}} \quad.
\end{equation}

	By assuming (like in Einstein's derivation), that at very high temperatures the stimulated emission is much more probable than spontaneous emission, (\ref{pnpm})
yields
\begin{equation}\label{pb}
	p_{n} B^{m}_{n}= p_{m} B^{n}_{m} \quad.
\end{equation}
 
Now, comparing (9) with the generalized Wien law (7), we obtain

\begin{equation}\label{abalfani}
	 {A^{n}_{m}\over B^{n}_{m}}=\alpha (\frac {N_r(t)}{N_{or}})^\frac {4}{3}\nu^{3}  \quad,
\end{equation}
and

\begin{equation}
E_m - E_n = (\frac {N_r(t)}{N_{or}})^\frac {1}{3} {h\nu} \quad. 
\end{equation}
Note that the time-dependent prefactors in the above relations appear as a consequence of the detailed balance 
between matter and radiation, that is, regardless of the 
specific gravitational theory 
for ``adiabatic'' photon creation.\footnote{Naturally, we are not 
advocating here that the fundamental constant $h$ is a time dependent 
quantity.} Finally, inserting relations (10)-(12) into (9) and fixing $\alpha$ taking the usual classical limit, we obtain 

\begin{equation}
\rho _{T}(\nu) = (\frac {N_r(t)}{N_{or}})^\frac {4}{3} \frac {8 \pi h}{c^{3}}
\frac {\nu ^{3}}  
{exp[(\frac {N_r(t)}{N_{or}})^\frac{1}{3} {\frac {h\nu}{kT}}]  - 1}   \quad.
\end{equation}
In the absence of creation ($N_{r}(t)=N_{or}$), the standard Planckian 
spectrum is recovered. In addition, since the
the exponential factor is time independent, this spectrum is not 
destroyed as the universe 
evolves. More important still, (13)
cannot, on experimental grounds, be distinguished from 
the blackbody spectrum at the present epoch when $T=T_o$ and $N_r(t_o)=
N_{or}$. Therefore, {\it models with photon creation can be compatible 
with the present day isotropy and spectral distribution of the CMBR}. 
This conclusion is extremely general. It does not depend either on the time 
dependence of the scale factor nor even on the 
form of $N_r(t)$, which in turn
must be determined by the specific photon creation theory. 
It is interesting that no reference has been made to the specific source of photons. The above derivation depends only on the new temperature law, or equivalently, on the existence of the generalized ``adiabatic'' invariant 
given by (3). This fact cannot be fortuitius; it must 
reflect (for photon creation) the same sort of 
universality contained in the  Planck distribution.
For consistency we observe that the equilibrium relations 
are recovered using the above spectrum. By introducing the 
variable 
$x=(\frac {N_r}
{N_{or}})^\frac {1}{3} \frac {h\nu}{kT}$, we obtain

\begin{equation}  
n_r (T) =\int_{0}^{\infty}
\frac {\rho_{T}(\nu) d\nu}
{(\frac {N_r(t)}
{N_{or}})^\frac {1}{3} 
h\nu} =bT^{3} \quad,
\end{equation}
and

\begin{equation} 
\rho_r (T) =\int_{0}^
{\infty} \rho_{T}(\nu )d\nu = 
aT^{4} \quad,
\end{equation}
where 
$b=\frac {0.244}
{{\hbar^
{3}} c^
{3}}$ 
and
$a=\frac {\pi^
{2} k^{4}}
{{15\hbar^
{3}} c^{3}}$, are 
the 
blackbody radiation 
constants\cite{MP 14}. 
 
To point out its intrinsically irreversible 
character, let us now 
clarify the concept of ``adiabatic'' photon 
creation in connection with the second law of 
thermodynamics. For a photon gas the Euler relation is 
  
\begin{equation}
\sigma_r=\frac{\rho_r + p_r}{n_rT}  \quad ,
\end{equation}
where $\sigma_r$
is the specific radiation entropy.  
In the homogeneous case, $\sigma_r = S_r/N_r$, where 
$S_r$ is the total photon
entropy. It thus follows that 
\begin{equation}
\dot \sigma_r=\frac {S_r}{N_r}
{(\frac {\dot S_r}
{S_r}-\frac{\dot N_r}{N_r})} \quad ,
\end{equation}
where a dot stands for time derivative. 
Since the form of adiabatic relations are 
preserved, (16) implies that 
$\dot \sigma_r=0$, and from (17)  

\begin{equation}
\frac{\dot S_r}{S_r} = \frac{\dot N_r}{N_r} 
\quad .
\end{equation}
Therefore, ``adiabatic'' creation means that the 
total entropy 
increases, as 
required by the second law of 
thermodynamics, however,
the specific entropy(per photon)
remains constant during the process. As a 
consequence, theories with photon creation must be
constrained by the second law of 
thermodynamics.\footnote{Using nonequilibrium 
relativistic thermodynamics
one may proof that the validity of the equilibrium  relations leads to 
$\dot \sigma_r=0$ and reciprocally. For decaying vacuum cosmologies see 
section 4 of Ref.\cite{JA 96}.} 
In this connection, we recall that the standard FRW model is consistent 
with the present CMBR spectrum but, it does not provide any explanation 
for the origin of the cosmological entropy. In principle, only 
a cosmology allowing entropy
production may be able to provide a definite solution to this problem.
\vskip 0.3cm
\noindent{\bf Conservation Versus Creation}
\vskip 0.3cm    
The present Planckian spectrum of the CMBR cannot 
be distinguished from (13), however, this does not 
mean that the same holds 
for moderate or high redshifts.  Using the scale factor-redshift relation, 
$R = R_o(1 + z)^{-1}$, the temperature law (1) becomes
 
\begin{equation}
T = T_o(1 + z)(\frac {N_r(t)}{N_{or}})^{\frac{1}{3}}  
\quad , 
\end{equation}
where $T_o$ is the present day value of $T$. This relation has some 
interesting physical consequences. Since
$N_r(t) \leq N_{or}$, it implies that  
universes with photon creation are, for 
any value of $z>0$, cooler than the
standard model. This prediction may indirectly be 
verified observing atomic or molecular transitions 
in absorbing clouds at large                                                                                                         redshifts. In this way, it provides a crucial test 
for models endowed with ``adiabatic'' photon production, which 
is acessible with present day 
technology\cite{JA 96,SO 94}. Qualitatively, (19) also explains why 
models with ``adiabatic'' photon creation solve the cosmological 
age problem which plagues the class of FRW models\cite{LGA 96}.
Since for a  given redshift $z$ the universe is cooler than in the 
standard model, more time is required                                                                                                                                                                                                                          to attain a fixed temperature scale in the early universe. 

In conclusion, we stress that photons injected in the spacetime with 
the ``normal'' distribution (13) cannot be responsible for the present 
observed distortions of the CMBR spectrum. Since this result follows 
naturally from the temperature law (1), cosmologies endowed with 
continuous photon creation are not disproved  by the blackbody nature 
of the CMBR.

\section*{Acknowledgments}

It is a pleasure to thank R. Brandenberger, A. Maia 
and V. Zanchin for a critical reading of the 
manuscript. Many thanks are also due to R. Abramo, R. Moessner and M. Parry 
for the permanent encouragment. This work was 
partially supported by the Conselho 
Nacional de Desenvolvimento Cient\'{i}fico e Tecnol\'{o}gico - CNPq (Brazilian
Research Agency), and by the US Department of Energy under grant 
DE-F602-91ER40688, Task A.

\end{document}